\documentstyle[twocolumn,aps,epsfig,floats]{revtex}
\draft 

\begin{document} 
\twocolumn[\hsize\textwidth\columnwidth\hsize\csname @twocolumnfalse\endcsname
\title{
The local magnetic moments and hyperfine magnetic
fields in disordered metal-metalloid alloys
}
\author{
A.K.Arzhnikov and L.V.Dobysheva}

\address{
Physical-Technical Institute, Ural Branch of Russian Academy of Sciences, \\
Kirov str. 132, Izhevsk 426001, Russia
}
\maketitle

\begin{abstract}
The local magnetic moments and hyperfine magnetic fields (HFF) in the 
ordered alloys $Fe_{15}Sn$ and $Fe_{15}Si$ are calculated with 
the first-principles full-potential linear augmented plane wave (FP LAPW) 
method. The results are compared with the experimental data on $Fe-M$ 
($M=Si$, $Sn$) disordered alloys at small metalloid concentration. 
The relaxation of the lattice around the impurity and its influence on 
the quantities under consideration are studied. The mechanism of the 
local magnetic moment formation is described. It is proved that the main 
distinction between these alloys is connected with the different lattice 
parameters. Three contributions to the HFF are discussed: the contributions
of the core and valence electron polarization to the Fermi-contact part, 
and the contibution from the orbital magnetic moment. 
\end{abstract}

]
\narrowtext  

     This paper deals with the low-concentration dependences of the
local magnetic moments and hyperfine magnetic fields (HFF) at the $Fe$
nuclei for the most typical metal-metalloid alloys $Fe_{1-x}Si_x$ and
$Fe_{1-x}Sn_x$.

\begin{figure}[bt]  
\epsfig{file=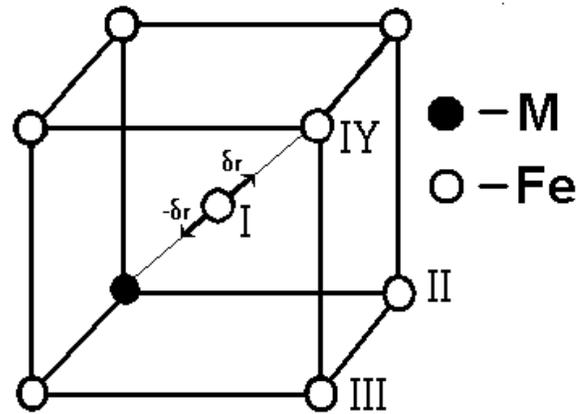,width=8.5cm}
  \caption{Unit cell of the ordered alloy  $Fe_{15}M$
}
 \label{cell}
\end{figure}

\begin{figure}[t]  
\epsfig{file=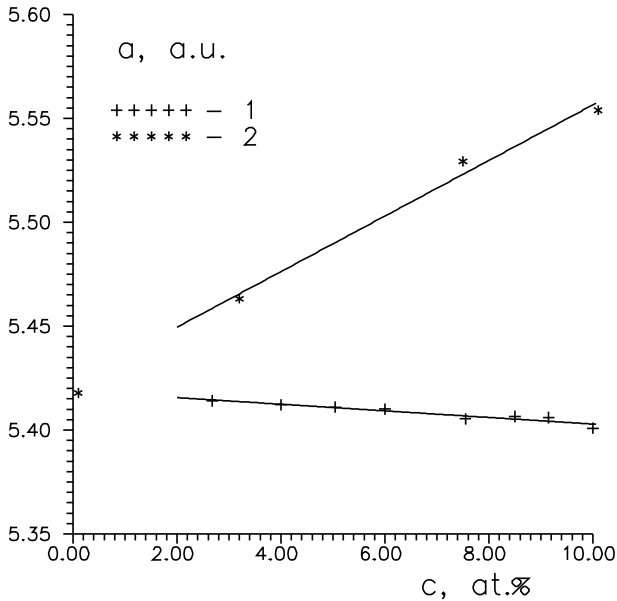,width=7.cm}
\vspace{1cm}
  \caption{Concentrational dependence of the BCC lattice parameter 
for the disordered alloys $Fe-Si$ (1) and $Fe-Sn$ (2)
}
  \label{lattice} 
\end{figure}

The first-principles calculations were performed for the ordered
alloys $Fe_{15}Sn$ and $Fe_{15}Si$ with lattices 
obtained by substituting a metalloid atom for one of the $Fe$
atoms in the extended BCC cell of 16 atoms. The unit cell 
contains 4 non-equivalent $Fe$ atoms located at different distances
from the metalloid atom (Fig.~\ref{cell}). 
Such a model of the alloy was chosen in accordance with the 
experimental data \cite{Els4,Els5} which show that the BCC structure
of the substitution alloys is retained up to $30 at.\%$ of metalloid.
Fig.~\ref{lattice} displays the BCC structure parameter taken for $Fe-Si$ 
from \cite{Els5,Farq,Rich}  and for $Fe-Sn$ from \cite{JETPh}.
The self-consistent band structure calculations were performed using
the full-potential linearized augmented plane waves (FLAPW) method with
the WIEN97 program package \cite{Wien1,Wien2}.

     The experimental lattice constants taken for the impurity
concentration of $6.25 at.\%$ (Fig.~\ref{lattice}) and extrapolated to
the zero temperature were used (for $Fe_{15}Sn$ $a_{Sn}=10.9924 a.u.$, for
$Fe_{15}Si$ $a_{Si}=10.7926 a.u.$). To illustrate the lattice parameter
effect, we carried out the calculations for $Fe_{15}Sn$ with $a_{Fe}=10.8114
a.u.$ corresponding to pure $Fe$. For $Fe_{15}Sn$ and $Fe_{15}Si$
the relaxation was studied by shifting the iron atoms $FeI$ closest to
the metalloid atom along the main diagonal of the cube.  The minimum
total energy was obtained for a shift $\delta r=0.008 a_{Sn} \sqrt{3}$ 
for tin
and $\delta r=-0.001 a_{Si} \sqrt{3}$ for silicon, which corresponds to the
experimental tendency of the lattice expansion/contraction in the case
of $Sn/Si$ (Fig.~\ref{lattice}). This indicates that a slightly
distorted BCC structure should occur in reality (see Fig.~\ref{cell}).

\begin{figure}[t]  
\vspace{6.5cm}
\epsfig{file=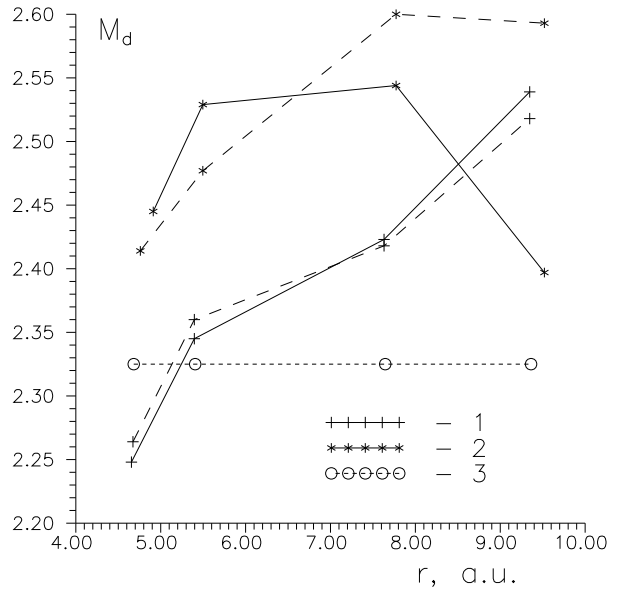,width=1.1cm}
  \caption{Local magnetic moments ($\mu_B$) (integrated spin density 
of the d-like electrons) at $Fe$ atoms in four
non-equivalent positions, 1 - $Fe_{15}Si$, 2 - $Fe_{15}Sn$, 3 - pure
$Fe$.
Dashed/solid lines are guidelines for values in non-relaxed/relaxed
lattice.
}
   \label{moments}
\end{figure}

     Hereafter the "local magnetic moment" means the integrated
spin density of the $d$-like electrons in the muffin-tin sphere $M_d$.
Fig.~\ref{moments} shows the local magnetic moments at the $Fe$ atoms
as a function of the distance to the metalloid atom in the ordered alloys
$Fe_{15}M$ ($M=Sn, Si$) for the non-relaxed and relaxed structure.
The $M_d$ magnitude is governed by two opposite factors. The first is
connected with the effective overlap of the $d$-like wave functions
that decreases with the interatomic distance or impurity amount.
The second deals with the $d$-band flattening with the increase of the
$s-d$ hybridization at an iron site due to the potential distortion 
by impurity. The local character of the $s-d$ hybridization results in
the fact that this effect takes place only near the impurity. The
distinctions in the local magnetic moments at the equivalent positions
between the alloys with $Si$ and $Sn$ are a result of different
inter-atomic distances. To prove this, the analysis of the $Fe_{15}Sn$
alloy with the lattice parameter of the $Fe_{15}Si$ system was
conducted. The $M_d$ values in alloys with $Si$ and $Sn$ were close if
the lattice parameters were taken equal. The $M_d$ values increase with
the lattice parameter, which is also confirmed by the calculations for
pure iron with different lattice parameters. One can see
(Fig.~\ref{moments}) that decreasing the distance from $FeIY$ to the
nearest neighbours in the relaxed system $Fe_{15}Sn$ results in a
decrease of the magnetic moment which approaches that of pure iron. We
should note that the distance to the farther spheres determined by the
lattice parameter remains completely unchanged or, for some spheres,
unchanged on the average, thus, the $FeIY$ magnetic moment is governed
mainly by the distance to the nearest neighbours.  The difference in
the total magnetization for the systems with $Si$ and $Sn$ corresponds
to the experimental value at the concentration $6.25 at.\%$.

     The electron-polarization contribution to the HFF was calculated
by the standard procedure of electron spin-density integration with the
relativistic effects taken into account \cite{Akai}. This contribution
includes two terms: the polarization of the inner-levels electrons
("core" electrons) in the nucleus region $H_{cor}$ and that of the valence
electrons $H_{val}$. The polarization of the core electrons follows the
relation $H_{cor}=\gamma M_d$ closely.  Usually, $H_{cor}$ is 
considered to be proportional to the total magnetization, which is
much worse. The factor $\gamma$ depends on
the type of the exchange-correlation potential. On the whole, the
calculations were done with the generalized gradient approximation
(GGA) of the potential \cite{GGA}, this gave $\gamma=-123 kG/\mu_B$.
The calculations with the local-density approximation (LDA) of the
potential \cite{LDA} change $\gamma$ to $-112 kG/\mu_B$. However, within
the approximation chosen, $\gamma$ keeps constant with an accuracy of
one percent for different alloys (with $Si$ or $Sn$), different
lattices (BCC, hexagonal), different lattice parameters.  The
dependence of $H_{val}$ on the distance to the impurity has an
oscillatory shape and resembles the Ruderman-Kittel-Kasuya-Yosida 
(RKKY) polarization. As shown in
\cite{FTT}, the simple function $\cos(2K_F r)/r^3$ obtained in the 
model of free-electron polarization reflects the main features of the
RKKY polarization in these alloys, i.e. the period, amplitude, phase;
but it does not take into consideration the spatial distribution of the
$s-d$ exchange interaction and inhomogeneity of the valence-electron
density, and so does not allow a quantitative analysis.

     A large contribution to the HFF is due to the orbital magnetic
moment $M_{orb}$. In pure iron, $M_{orb}$ resulting from the reduction
of the Hamiltonian symmetry by the spin-orbit correction is of $0.08
\mu_B$. With the metalloid impurity inclusion, the cubic symmetry of 
the crystal potential is also violated, which gives an additional
defreezing of the orbital magnetic moment, and its growth leads to
a decrease in the HFF absolute value. This is corroborated by almost
equal experimental differences $H_0-H_1\approx 20 kG$ for both
$Fe_{100-c}Si_c$ \cite{Els3} and $Fe_{100-c}Sn_c$ \cite{Els1} ($H_0$ is
the HFF at the nucleus of $Fe$ without metalloid atoms in the nearest
environment, and $H_1$ is that with one metalloid atom in the nearest
environment). This difference could not be explained by either
change of the core electron polarization ($H_{cor}$) due to the 
magnetic moment change (Fig.~\ref{moments}), or the change in the 
RKKY polarization ($H_{val}$)
as was shown in \cite{FTT}. We would like to remind that slightly
increasing the orbital magnetic moment by $\approx 0.06 \mu_B$ should
give a decrease of $20 kG$ in the HFF absolute value \cite{Ebert}.
Though present first-principles calculations do not give exactly the
experimental values of the orbital magnetic moment, our computations
\cite{IEM} showed that the orbital magnetic moment and its contribution 
to the HFF increase when an impurity appears in the nearest environment
of an iron atom.

This work was supported by Russian Fund for Basic Research.


\end{document}